# Comment: Bayesian Checking of the Second Levels of Hierarchical Models

**M. Evans**

*Abstract.* We discuss the methods of Evans and Moshonov [*Bayesian Analysis* **1** (2006) 893–914, *Bayesian Statistics and Its Applications* (2007) 145–159] concerning checking for prior-data conflict and their relevance to the method proposed in this paper.

## 1. INTRODUCTION

This is an interesting paper dealing with an important topic. It is a logical continuation of the contributions found in Bayarri and Berger (2000). In particular, it continues the emphasis on avoiding the "double use of the data" and this is an important point that we agree with.

While it seems intuitively clear what "double use of the data" means, it would be nice to have a precise definition as the phrase seems to be used a bit too freely by some at times, at least in our view. Intuitively, in model checking, this would seem to be the situation where the fitted model depends on a particular aspect of the data and then the model is checked by comparing the same aspect of the data with the fitted model. On the other hand, we have seen assertions that a "double use of the data" is being made in situations like computing a posterior (the first use) and then (the second use) computing a characteristic of that distribution like a mode or hpd region. While in some technical sense this seems like using the data twice, there does not seem to be anything wrong with it, at least to us. Rather than giving a definition, this paper, like Bayarri and Berger (2000) and Robins, van der Vaart and Ventura (2000), points to a negative consequence of double use of the data, in terms of the lack of uniformity of $p$-values. Perhaps the factorization in Section 2 of this discussion gives a general method of ensuring that components of the total information available to a statistician for an analysis are used appropriately, and so gives a general characterization for avoidance of "double use of the information."

This paper assumes a default or "objective" prior on the last level of a hierarchically specified prior. In general this will result in an improper prior. Part of the motivation for this seems to be that "model checking with informative priors cannot separate inadequacy of the prior from inadequacy of the model" and so the methodology proposed by Box (1980), which is based on proper priors, is not used. We disagree with the quoted statement. The methods discussed in Evans and Moshonov (2006, 2007) are a modification of Box's approach and are motivated precisely by the need to separate the two kinds of inadequacies in the context of proper, informative priors which, as they should, represent subjective beliefs. We briefly outline this approach in Section 2. Also, Evans and Moshonov (2006) includes methodology for checking the second level of a hierarchical model based on a factorization of the full information. We discuss this in Section 3 and show that this methodology is also applicable when the first level is improper.

While we agree with the necessity to consider improper priors as part of a general theory of statistics, it is difficult for us to accept these as a basis from which statistical theory is built. It is our opinion that the core of statistics is represented by the proper prior context. As such, we feel that what is done outside of this core should be highly influenced, if not directed, by the central theory with proper priors. So our discussion reflects this and considers the implications for the situation discussed in this paper.

*M. Evans is Professor, Department of Statistics, University of Toronto, Toronto, Ontario M5S 3G3, Canada e-mail: mevans@utstat.utoronto.ca.*

For us checking the sampling model and the prior are important parts of a statistical analysis. A common complaint concerning the prior is that it is subjective, as it represents someone's personal beliefs about the true value of $\theta$. A common retort is that the sampling model is also subjective as it represents someone's belief that the true distribution is in this class, that is, it was someone's subjective choice. Of course, both these statements are correct as there is typically little "objective" about either choice. From another point of view, the fact that these choices are subjective is a good thing because they are (hopefully) informed choices and that should lead to better statistical analyses than if we made these choices arbitrarily, or based on convention. For us the way to reconcile the debate between objective and subjective is through checking that these ingredients make sense in light of what we know to be truly objective (at least if it is collected correctly), namely, the data. Others argue that no such checks should be made, as they lead us to be incoherent. There is a wide diversity of opinion on these matters and we certainly acknowledge value in various points of view.

## 2. FACTORING THE FULL INFORMATION

Suppose we have prescribed a sampling model $\{P_\theta : \theta \in \Theta\}$, a proper prior $\Pi$, and have observed the data $x$. The sampling model and prior combine to give the joint model $P_\theta \times \Pi$ for $(x, \theta)$. We will suppose that this joint model and the observed data comprise the full information available to the analyst. We are not saying that further information may not be available in an analysis, but we will restrict our discussion to the situation where this is all we have. Further, denote the prior predictive measure by $M(B) = \int_\Theta P_\theta(B)\Pi(d\theta)$; for statistics $T$ and $U \circ T$ on the sample space let $M_T(\cdot|U \circ T)$ denote the conditional prior predictive distribution of $T$ given $U \circ T$, and $\Pi(\cdot|x)$ denote the posterior of $\theta$.

In Box's approach to model checking, the observed value of $x$ is compared with $M$ to see if there is model failure, that is, we check to see if $x$ is a surprising value from $M$. There would appear to be an illogicality involved in this, however, as we know, at least in the subjective Bayesian context, that $x$ was not generated from $M$. If our assertion was that $x$ was generated from $M$, perhaps as a random effects model, then it would make sense to check $x$ against $M$, as this is an assertion about the underlying data generating mechanism. It is clearly more appropriate, in Bayesian context, however, to see if $x$ is not surprising for at least one of the distributions in $\{P_\theta : \theta \in \Theta\}$, that is, check $x$ against what we are asserting is the data generating mechanism—the sampling model.

As discussed in Evans and Moshonov (2006), there are two possibilities for failure in the Bayesian formulation: the sampling model may fail by $x$ being surprising for each distribution in the sampling model or, if the sampling model does not fail, the prior may conflict with the data by placing the bulk of its mass on those distributions in the sampling model for which the data is surprising. Note that it only makes sense to talk about prior-data conflict if the sampling model does not fail. Logically, checking the sampling model precedes checking for prior-data conflict.

How then should we check for prior-data conflict? Intuitively this arises when the effective supports of the likelihood and the prior do not overlap. As discussed in Evans and Moshonov (2006), however, the clearest approach to measuring this conflict comes from asking if the observed likelihood is a surprising value from its prior predictive distribution. Given that the likelihood map is minimal sufficient, this is equivalent to asking if the observed value $T(x)$ of a minimal sufficient statistic $T$ is surprising from its marginal prior predictive $M_T$. Further consideration shows that $T(x)$ can be surprising simply because some value $U(T(x))$ is surprising where $U \circ T$ is ancillary. When such ancillaries exist, this leads to comparing $T(x)$ to $M_T(\cdot|U \circ T)$ where $U \circ T$ is a maximal ancillary, as this conditioning removes the maximal amount of ancillary variation. Ancillary variation is clearly not relevant to assessing prior-data conflict as it does not depend on the parameter. Further, there is nothing to prevent us from using some function $S(T)$, and comparing its observed value to the distribution $M_{S(T)}(\cdot|U \circ T)$, to check for prior-data conflict. Of course, $S$ has to be chosen sensibly if we are going to make a meaningful check.

This approach leads to the following factorization of the joint distribution:

$$P_\theta \times \Pi$$

(1)
$$= P(\cdot|T) \times P_{U \circ T} \times M_T(\cdot|U \circ T) \times \Pi(\cdot|x),$$

where $P(\cdot|T)$ is the conditional distribution of the data given the minimal sufficient statistic $T$, and so does not involve $\theta$, and $P_{U \circ T}$ is the marginal distribution of $P_{U \circ T}$ which is also free of $\theta$. Each of the components in (1) plays a separate role in a statistical analysis. $P(\cdot|T)$ and $P_{U \circ T}$ are available for checking the sampling model, $M_T(\cdot|U \circ T)$ is available for checking for prior-data conflict and $\Pi(\cdot|x)$ [which really only depends on the data through $T(x)$] is for inference about $\theta$. We see that $M = P(\cdot|T) \times P_{U \circ T} \times M_T(\cdot|U \circ T)$, which explains how this is a modification of Box's approach and it shows how to check for inadequacies in the prior as well as the sampling model.

It is our claim that effectively (1) shows us how to proceed to avoid double use of the information and, as such, avoid double use of the data. Of course, as mentioned in the paper, it may be difficult, with complicated models, to determine $P(\cdot|T)$ or $P_{U \circ T}$ in meaningful ways. Accordingly, it seems reasonable to weaken this requirement in such contexts to having this hold asymptotically in some sense. For example, a chi-squared goodness-of-fit test is asymptotically ancillary.

In the context of an improper prior that leads to a proper posterior, then (1) is still available but now the factor $M_T(\cdot|U \circ T)$ is not a probability measure and so it is not clear how we would check for prior-data conflict. As discussed in Evans and Moshonov (2006, 2007), a partial characterization of a noninformative prior is that it would never lead to evidence of a prior-data conflict existing no matter what data is obtained. Thus the choice of an improper prior is an assertion that this choice avoids such a conflict. Noninformative sequences of priors are also discussed in Evans and Moshonov (2006, 2007) and these can provide a way to justify such a statement for a particular improper prior. In any case, the choice of an improper prior should not in any way change the role of the remaining factors if we follow the principle that the proper case is central. Although we do not have a formal proof, it would seem that the methods discussed in Bayarri and Berger (2000) will satisfy this asymptotically.

Further, any $p$-values computed according to this factorization will have the necessary uniform properties when assessed against the appropriate measures. For example, if $p(t) = M_T(h(T) > h(t))$ is a $p$-value for checking for prior-data conflict with no ancillary, then $p(T)$ will be uniformly distributed, at least in the continuous case, when $T \sim M_T$.

## 3. HIERARCHICAL MODELS

In Evans and Moshonov (2006, 2007) methods are discussed for checking hierarchically specified priors for $\theta = (\theta_1, \theta_2) \in \Theta_1 \times \Theta_2$, that is, we specify priors $\Pi_1$ and $\Pi_2$ so that $\Pi(d(\theta_1, \theta_2)) = \Pi_2(d\theta_2|\theta_1) \times \Pi_1(d\theta_1)$. In such situations we would like to check the individual components of the prior separately, as this gives us more information about a prior-data conflict when this occurs. For example, it may be that $\Pi_1$ conflicts but $\Pi_2$ does not.

We distinguish two different situations. First, the parameters $\theta_1$ and $\theta_2$ may both be part of the likelihood function and second, only $\theta_2$ is part of the likelihood function. The second situation corresponds to hierarchical models and $\theta_1$ is a hyperparameter. Methods are presented in Evans and Moshonov (2006, 2007) for both of these situations, but we only discuss hierarchical models here.

With proper priors we have the prior $\Pi_2^*(d\theta_2) = \int_{\Theta_1} \Pi_2(d\theta_2|\theta_1)\Pi_1(d\theta_1)$ for the model parameter and the methods of Section 2, based on the minimal statistic $T$ for the model $\{P_{\theta_2} : \theta_2 \in \Omega_2\}$, are available to check whether or not $\Pi_2^*$ conflicts with the data. While this check is available, Evans and Moshonov (2006) develop a factorization that is appropriate for checking the components, such as the second level $\Pi_2(\cdot|\theta_1)$, of a hierarchical model.

To simplify the presentation of this, we will suppose there are no relevant ancillaries for $\{P_{\theta_2} : \theta_2 \in \Omega_2\}$ based on $T$, but note that these can be incorporated as well. We can formally generate another model for $x$ from the joint distribution, namely, via

$$\begin{aligned}
M_{\theta_1}(dx) &= \int_{\Omega_2} P_{\theta_2}(dx)\Pi_2(d\theta_2|\theta_1) \\
&= P(dx|T)(t) \int_{\Omega_2} P_{T\theta_2}(dt)\Pi_2(d\theta_2|\theta_1) \\
&= P(dx|T)(t) \times M_{T\theta_1}(dt).
\end{aligned}$$

This model is only formal, as, indeed, our model indicates that $x$ was not generated via $M_{\theta_1}$, for some value of $\theta_1$. Here $M_{\theta_1}$ is the conditional prior predictive distribution for $x$ given $\theta_1$ and $M_{T\theta_1}$ is the conditional prior predictive distribution for $T$ given $\theta_1$. Note that when $\Pi_2(\cdot|\theta_1)$ is proper, as in the paper, then $M_{\theta_1}$ and $M_{T\theta_1}$ are also proper.

Let $V(T)$ be a minimal sufficient statistic for the formal model for $T$ given by $\{M_{T\theta_1} : \theta_1 \in \Omega_1\}$. We can factor $M_{T\theta_1}$ as $M_T(\cdot|V) \times M_{V\theta_1}$, where $M_T(\cdot|V)$ is the conditional prior predictive distribution of $T$ given $V$, and $M_{V\theta_1}$ is the conditional prior predictive distribution of $V$ given $\theta_1$. Then the joint distribution of $(\theta_1, x)$ can be factored as

(2) $\quad P(\cdot|T) \times M_T(\cdot|V) \times M_V \times \Pi_1(\cdot|V),$

where $M_V$ is the prior predictive distribution of $V$ and $\Pi_1(\cdot|V)$ is the posterior distribution of $\theta_1$.

Consider how each of the factors in (2) is to be used. First $P(\cdot|T)$ is available for checking the basic sampling model $\{P_{\theta_2} : \theta_2 \in \Omega_2\}$. If no evidence is found against $\{P_{\theta_2} : \theta_2 \in \Omega_2\}$, we can proceed to check the formal model $\{M_{T\theta_1} : \theta_1 \in \Omega_1\}$ for $T$ using $M_T(\cdot|V)$ and note that this does not depend on $\Pi_1$. Note also that $M_T(\cdot|V)$ is proper whenever $\Pi_2(\cdot|\theta_1)$ is proper for each value of $\theta_1$. If evidence is found against this model, then, because we have accepted the sampling model, and so consequently the model $\{P_{T\theta_2} : \theta_2 \in \Omega_2\}$ for $T$, this must occur because of a conflict between the observed value $T(x)$ and $\Pi_2$. So a check of the formal model $\{M_{T\theta_1} : \theta_1 \in \Omega_1\}$ using $M_T(\cdot|V)$ is a check for prior-data conflict with $\Pi_2$. Note that this check proceeds exactly as in the simpler situation described in Section 2. If we find no evidence against $\{M_{T\theta_1} : \theta_1 \in \Omega_1\}$, then we can check for a conflict with $\Pi_1$ using $M_V$. Finally, if there is no conflict with $\Pi_1$, then $\Pi_1(\cdot|V)$ is available for inference about $\theta_1$. Of course, if there is no conflict with $\Pi_1$ and $\Pi_2$, then we can also make inference about the parameter of interest $\theta_2$.

The model $\{M_{V\theta_1} : \theta_1 \in \Omega_1\}$ may have ancillaries. Let $W \circ V$ be such a maximal ancillary. We then have that $M_V$ factors as $M_V = M_{W \circ V} \times M_V(\cdot|W \circ V)$ so that (2) becomes

(3) $\quad P(\cdot|T) \times M_T(\cdot|V)$
$\quad\quad \times M_{W \circ V} \times M_V(\cdot|W \circ V) \times \Pi_1(\cdot|V).$

In this case, given that we have accepted the sampling model, the factor $M_{W \circ V}$ is available for checking for prior-data conflict with $\Pi_2$, and $M_V(\cdot|W \circ V)$ is the appropriate factor for checking $\Pi_1$. The justification for this is exactly as in the simple case discussed in Section 2.

Note that in (3), the only distribution that will necessarily be improper when $\Pi_1$ is improper, is $M_V(\cdot|W \circ V)$. The measure $M_V(\cdot|W \circ V)$ is to be used only in the check for $\Pi_1$. Therefore, the choice of an improper $\Pi_1$ is really an assertion that this prior will never conflict with the data. Irrespective of whether or not $\Pi_1$ is improper, the factors $M_T(\cdot|V)$ and $M_{W \circ V}$ are available to check for prior-data conflict with $\Pi_2$, when it is proper.

We consider the implementation of this approach in the normal-normal hierarchical model presented in the paper.

EXAMPLE (*Normal–normal hierarchical model*). We first consider a simpler model. In particular, we assume that the known $\sigma_i^2$ are all equal to $\sigma^2$ and that we have balance, namely, $n_1 = \cdots = n_I = n$. For this problem we have that $T(x) = (\bar{x}_1, \ldots, \bar{x}_I)' \sim N_I(\theta, (\sigma^2/n)I)$ and here $\theta$ is the model parameter (corresponding to $\theta_2$ in our parameterization of a hierarchical model above). Therefore, according to our factorization, we check the sampling model using $P(\cdot|T)$, which is effectively the distribution of the residuals.

Now

$$(\bar{x}_1, \ldots, \bar{x}_I)' = (\theta_1, \ldots, \theta_I)' + (\sigma/\sqrt{n})(z_1, \ldots, z_I)'$$

where the $z_i$ are i.i.d. $N(0,1)$ and, from the second level, $(\theta_1, \ldots, \theta_I)' \sim N_I(\mu 1, \tau^2 I)$, independent of $(z_1, \ldots, z_I)'$. Thus $(\mu, \tau^2)$ is the hyperparameter (corresponding to $\theta_1$ in our parameterization of a hierarchical model above). This implies that $M_{T(\mu,\tau^2)}$ is given by $(\bar{x}_1, \ldots, \bar{x}_I)' \sim N_I(\mu 1, (\tau^2 + \sigma^2/n)I)$. It is then easy to see that $V(\bar{x}_1, \ldots, \bar{x}_I) = (\sum_{i=1}^I \bar{x}_i, \sum_{i=1}^I \bar{x}_i^2)$ is a minimal sufficient statistic for the model $\{M_{T(\mu,\tau^2)} : \mu \in R^1, \tau^2 > 0\}$. Note also that $V$ is a complete minimal sufficient statistic so there are no relevant ancillaries $W$ that we need consider for the check for the second level.

To determine $M_T(\cdot|V)$ we need the conditional distribution of $(\bar{x}_1, \ldots, \bar{x}_I)'$ given $(\sum_{i=1}^I \bar{x}_i, \sum_{i=1}^I \bar{x}_i^2)$. This is clearly uniform on the sphere of squared radius $\sum_{i=1}^I \bar{x}_i^2$ lying in the hyperplane of $R^I$ given by $\{(y_1, \ldots, y_I)' : \sum_{i=1}^I y_i = \sum_{i=1}^I \bar{x}_i\}$. We can simulate from this distribution by generating $v_1, \ldots, v_{I-1}$ i.i.d. $N(0,1)$, putting $u_i = v_i/(\sum_{i=1}^{I-1} v_i^2)^{1/2}$ and

$$(y_1, \ldots, y_I)' = (\bar{x}_1, \ldots, \bar{x}_I)' + A(u_1, \ldots, u_{I-1})'$$

where $A \in R^{I \times (I-1)}$ is such that the matrix $(1/\sqrt{I}\ A)$ is orthogonal. Then for any particular discrepancy statistic, we can compute an appropriate $p$-value via simulation.

The above analysis also applies when the $\sigma_i^2/n_i$ are all equal. When they are not equal the analysis is more complicated, as the form of $V$ depends on which ones are equal. Further, it is not a complete minimal sufficient statistic and so there are relevant ancillaries.

Based on the factorization (3) we feel that $M_T(\cdot|V)$ and $M_{W \circ V}$ are appropriate distributions for computing $p$-values to assess the second level for a hierarchical model. Further, the uniformity of the corresponding $p$-values should be assessed against these distributions and this does not require that $\Pi_1$ be improper.

It is difficult to compare our approach with the proposal in the paper, but we note that it has the distinct advantage of not involving the prior for the first level. For our check on the second level we need say nothing about the prior for the first level and it can be improper. The intuition for this lies with conditioning on $V$, which completely removes the effect of $\Pi_1$ on the prior predictive for $T$, and the fact that $\Pi_2$ induces the ancillary $W \circ V$. Therefore any conflict that is found can only be due to $\Pi_2$. It may be that the method proposed in the paper will satisfy (3) in an asymptotic sense but we do not have a proof of this.

## 4. CONCLUSIONS

It is sometimes suggested that model checking is a somewhat informal process. Partly this is because models can fail in many ways and some of these may be more relevant in certain situations than others. It seems impossible then to come up with a methodology that will check for all of the possibilities simultaneously. So it seems reasonable to ask that we specify a set of checks that we think are relevant, prior to seeing the data, and then implement only these, rather than going on a hunting expedition for defects. A similar approach seems appropriate for checking for prior-data conflict.

While selection of the actual checks is perhaps somewhat informal, we do not believe that there is complete freedom in this. Some general principles must apply. The ill effects of double use of the data, as discussed in this paper and Bayarri and Berger (2000), provide a good example of the need for such principles.

In frequentist statistical theory, inference about parameters depends on the data only through the minimal sufficient statistic and, what is left over in the data (the residual), is available for model checking. Mixing these up would seem to correspond to an inappropriate statistical analysis. We believe this is equally applicable in Bayesian formulations.

Checking for prior-data conflict seems to sit between model checking and inference. While it depends on the minimal sufficient statistic, however, the factorization given by (1) indicates that it really is separate from model checking and inference as it involves a separate component of the full information as expressed by the joint distribution. In essence (1) prescribes how each component of the full information is to be used in a statistical analysis. If we mix these up, it would seem to us that we can expect illogical or incoherent behavior, for example, overly conservative $p$-values. Note that in a certain sense each component of (1) is independent of the others, as we could prescribe each probability measure separately and still end up with a valid joint distribution. Specification of each component of (1) is necessary and sufficient for the specification of a joint probability distribution for $(x, \theta)$.

Of course, this restriction could be weakened to requiring that a methodology only satisfy (1) in some asymptotic sense. The motivation for this would seem to arise from the complexity of some situations. Still, (1) can be implemented exactly with many models of considerable importance, so it is not just of theoretical relevance.

Similarly, we believe that (3) is the relevant factorization for model checking and checking for prior-data conflict in hierarchical models. From that perspective it would be important to see if the methods proposed in the paper satisfied this in some asymptotic sense. This would give us more confidence that these constituted an appropriate way to proceed in situations where they were felt to be necessary.

We also feel that our discussion of (3) shows that the choice of prior $\Pi_1$ for $\theta_1$ is irrelevant for checking $\Pi_2$ with hierarchical models. In particular, whether $\Pi_1$ is proper or improper, the check for $\Pi_2$ is the same and this is a satisfying result. This does not appear to be the case for the method proposed in the paper which depends, in particular, on which objective prior we use. Perhaps this effect disappears as the amount of data increases, but then the relevance of checking for prior-data conflict disappears too, as the effect of the prior on inference disappears, at least under reasonable regularity conditions.

Overall, our purpose here is to suggest that there is a principled approach to the question addressed in the paper. We are not saying that using the partial posterior approach is in some way incorrect. We do think, however, that it would be worth investigating to what extent the partial posterior approach satisfied (3).